\documentclass[lengthcheck,letterpaper,superscriptaddress,twocolum,prb,aps,showpacs]{revtex4}
	\usepackage{amsmath}
	\usepackage[ansinew]{inputenc}
	\usepackage{graphicx}
	\usepackage{color}
	\usepackage[colorlinks,citecolor=blue,linkcolor=blue,urlcolor=blue]{hyperref}

\begin{document}

\preprint{New version incluiding some changes by PGG (in magenta)}

\title{Ab-initio nanoplasmonics: The impact of atomic structure}

\author{Pu Zhang}
\affiliation{Departamento de F\'{\i}sica Te\'orica de la Materia Condensada and Condensed Matter Physics Center (IFIMAC), Universidad Aut\'onoma de Madrid, E-28049 Cantoblanco, Madrid, Spain}

\author{Johannes Feist}
\affiliation{Departamento de F\'{\i}sica Te\'orica de la Materia Condensada and Condensed Matter Physics Center (IFIMAC), Universidad Aut\'onoma de Madrid, E-28049 Cantoblanco, Madrid, Spain}

\author{Angel Rubio}
\affiliation{Nano-Bio Spectroscopy group, Universidad del Pa\'{\i}s Vasco UPV/EHU,
CFM CSIC-UPV/EHU-MPC and DIPC, Avenida de Tolosa 72, E-20018 Donostia, Spain}
\affiliation{ETSF Scientific Development Centre, E-20018 Donostia, Spain}

\author{Pablo Garc\'{\i}a-Gonz\'alez}
\email{pablo.garciagonzalez@uam.es}
\affiliation{Departamento de F\'{\i}sica Te\'orica de la Materia Condensada and Condensed Matter Physics Center (IFIMAC), Universidad Aut\'onoma de Madrid, E-28049 Cantoblanco, Madrid, Spain}
\affiliation{ETSF Scientific Development Centre, E-20018 Donostia, Spain}

\author{F.\;J. Garc\'{\i}a-Vidal}
\email{fj.garcia@uam.es}
\affiliation{Departamento de F\'{\i}sica Te\'orica de la Materia Condensada and Condensed Matter Physics Center (IFIMAC), Universidad Aut\'onoma de Madrid, E-28049 Cantoblanco, Madrid, Spain}

\date{\today}

\begin{abstract}
We present an ab-initio study of the hybridization of localized surface plasmons in a metal nanoparticle dimer. The atomic structure, which is often neglected in theoretical studies of quantum nanoplasmonics, has a strong impact on the optical absorption properties when sub-nanometric gaps between the nanoparticles are considered. We demonstrate that this influences the hybridization of optical resonances of the dimer, and leads to significantly smaller electric field enhancements as compared to the standard jellium model. In addition we show that the corrugation of the metal surface at a microscopic scale becomes as important as other well-known quantum corrections to the plasmonic response, implying that the atomic structure has to be taken into account to obtain quantitative predictions for realistic nanoplasmonic devices.
\end{abstract}


\maketitle

There is a growing interest in the development and implementation of nanoplasmonic devices such as nanosensors \cite{Anker08,Dinh10}, nanophotonic lasers~\cite{Noginov09,Oulton09, Berini11}, optoelectronic~\cite{Ward10,Galperin12} and light-harvesting~\cite{Schuller10,Aubry10} structures, and nanoantennas~\cite{Muhlschlegel05,Novotny11}. Therefore, it is essential to have theoretical techniques with sufficient predictive value for understanding the physical processes of light-matter interaction at the nanoscale. In this regime, the standard analysis of the plasmonic response to external electromagnetic (EM) fields using the classical macroscopic Maxwell equations must be undertaken with caution. Indeed, genuine quantum effects such as the nonlocal nature of the electron-density response, the inhomogeneity of the conduction-electron density, or the possibility of charge transfer by tunneling have to be considered~\cite{Halas11}. These effects can be incorporated into Maxwell equations in an approximate manner using, e.g., nonlocal dielectric functions \cite{Ruppin01,McMahon09,Raza11,David11,Fernandez-Dominguez12a,Fernandez-Dominguez12b,Christensen14} or the \textit{ad-hoc} inclusion of ''virtual'' dielectric materials~\cite{Esteban12, Ma12,Luo13}. While these semi-classical approximations have been successfully applied in many cases, they never achieve the precision provided by first-principle calculations.

A number of recent publications \cite{Zuloaga09,Esteban12,Marinica12,Stella13,Andersen13,Teperik13a} have treated the electronic response of plasmonic structures using state-of-the-art time-dependent density functional theory (TDDFT) \cite{Runge84,Marques12}. However, the ionic structure is typically neglected and replaced by a homogeneous jellium background or by an unstructured effective potential. Although this approximation is sometimes justified by the collective nature of plasmon excitations \cite{deHeer93,Brack93}, the charge oscillations associated with a localized surface plasmon (LSP) are mainly concentrated on the metal-vacuum interface. One may thus expect that the ionic structure in this region will have quantitative and even qualitative impact. Therefore, there is a need to address the influence of the atomic configuration in the plasmonic response at the nanoscale.

In this Communication we present ab-initio calculations including the atomic structure, in accordance with the current paradigm in computational condensed matter physics \cite{Onida02} and physical chemistry \cite{Morton11}. We study one of the canonical cases in nanoplasmonics: the hybridization of LSPs in a metallic-cluster dimer. This structure has received widespread attention from a theoretical perspective in the last years \cite{Nordlander04,Romero06,Zuloaga09,Marinica12}, and has recently been experimentally realized in nanodevices with sub-nanometric gaps \cite{Savage12,Scholl13}. Furthermore, this is one of the prototypic plasmonic systems where quantum effects are relevant \cite{Tame13}. First, the coexistence of different natural length scales requires taking into account the non-locality of the response to EM fields. Second, the hybridization will depend very sensitively on the spacing between the effective surfaces of the clusters, which is determined by the amount of electron-density spill-out at the metal-vacuum interfaces \cite{Stella13,Teperik13b}. Finally, if the distance between the clusters is small enough, incident EM radiation can establish an alternating tunnel current between the two clusters, which greatly affects the electric field enhancement (EFE) in the interstitial region \cite{Zuloaga09,Marinica12}.

\begin{figure}[htbp] 
\begin{center}
\includegraphics[width=\columnwidth]{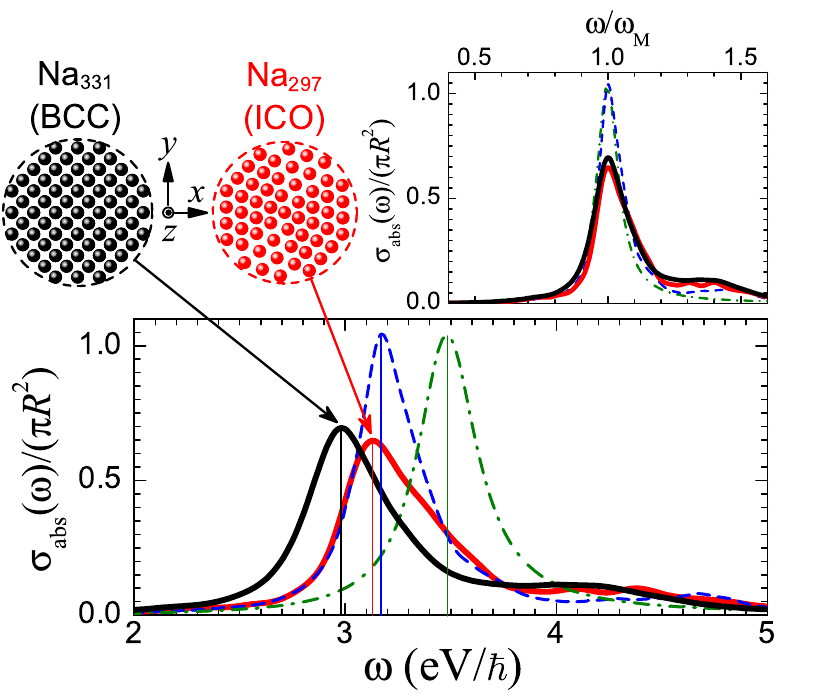}
\caption{(Color online). Main panel: normalized TDDFT optical absorption cross section for isolated Na clusters. Thick black line: atomic BCC arrangement for a $\mathrm{Na}_{331}$ cluster, thick red line: atomic ICO structure for a $\mathrm{Na}_{297}$ nanoparticle, dashed blue line: jellium model. The classical local-optics results (dashed-dotted green line) are included as well. The latter has been artificially broadened in order to fit the width of the main absorption peak of the jellium cluster. The main-resonance energies, $\hbar \omega_\mathrm{M}$, are: $2.98~$eV (BCC), $3.13~$eV (ICO), $3.17~$eV (jellium), and $3.49~$eV (classical local-optics). In the upper left panel we present pictures of the two atomistic arrangements (BCC and ICO). Upper right panel: Optical absorption cross sections for the four structures analyzed in the main panel but in each case the frequency is normalized to its respective resonance frequency 
$\omega_\mathrm{M}$.}
\label{fig1}
\end{center}
\end{figure}

Specifically, we analyze the optical response of nanodimers which are made of sodium clusters, although the conclusions can be extended with the appropriate modifications to any other metallic dimer, e.g., gold or silver. Since the aim of the present Letter is to analyze the impact of the atomic structure \textit{per se}, in what follows we will work with nanodimers in which the underlying ionic structure of the two clusters is maintained when the distance between them is varied \cite{Varas14}. We consider Na clusters with two different atomic arrangements: i) Na clusters of $331$ atoms that are arranged symmetrically around a central atom on a BCC lattice whose constant is set to the bulk Na experimental value ($a = 0.423~$nm) and ii) Na clusters of $297$ atoms exhibiting an icosahedral (ICO) symmetry \cite{Cambridge}. This geometry has been found to be the most stable configuration after optimization of the ionic positions in small clusters of alkali atoms  \cite{Noya07}. As can be seen in the upper left panel of \autoref{fig1}, both atomic clusters are almost spherical, the corresponding effective diameters being $2R_\mathrm{BCC}=2.88~$nm and $2R_\mathrm{ICO}\simeq2.71~$nm, respectively. The two atomistic descriptions will be compared against the jellium model, where the positive background is a homogeneous sphere of diameter $2R_\mathrm{jel}= 2.88~$nm, that is, the one corresponding to the $331$-atoms cluster. For each dimer, the separation between the clusters is defined as $d = b - 2R$ for both the atomic and jellium cases, where $b$ is the distance between the central atoms of each cluster or between the centers of the jellium spheres. Finally, only the $3s$ conduction electrons are explicitly included in the calculation by using standard norm-conserving Troullier-Martins pseudopotentials \cite{Trouiller91}.

Due to the nanometric size of the system, retardation effects can be safely ignored. The optical response can thus be evaluated using TDDFT under the adiabatic local density approximation, which is appropriate for simple metals such as Na~\cite{Rubio96,Vasiliev99,Joswig08}. To solve the TDDFT equations we follow a time-propagation/real-space prescription \cite{Yabana96} as implemented in the OCTOPUS package \cite{Marques03, Castro06}. According to this recipe, the ground-state electron system is perturbed at $t = 0$ by a delta-kick electric field $\mathbf{E}(\mathbf{r},t) = (\hbar \kappa_0 / e) \delta ( t )\mathbf{e}_x$, where $e$ is the absolute value of the electron charge and  $\kappa_0$ is small enough ($\kappa_0=0.005$ a.u.) to ensure a linear response by the electrons. As a result, the Kohn-Sham (KS) wavefunctions at $t=0^+$ are $\psi_i(\mathbf{r},0^+) = \exp(i\kappa_0 x)\psi_i(\mathbf{r})$, where $\psi_i(\mathbf{r})$ are the unperturbed ground-state KS orbitals. After this initial step, the time-dependent KS equations are solved in a real-domain representation and, in particular, the induced time-dependent electron density $\delta n(\mathbf{r},t)$ is obtained. Its frequency representation is thus given by
\begin{equation}
\delta n(\mathbf{r},\omega) = \int_{0}^{T_{\mathrm{max}}} \delta n(\mathbf{r},t) e^{(i\omega-\gamma) t} dt \; ,
\label{Eq1}
\end{equation}
where $T_{\mathrm{max}}$ is the total propagation time and $\gamma$ is a damping frequency which simulates non-electronic losses. In our calculations we have used $\gamma = 0.10~\textrm{eV}/\hbar$, which accounts for the linewidth of the absorption spectra of single Na clusters \cite{Yannouleas93}. Therefore, $(E_0 e / \hbar \kappa_0) \delta n(\mathbf{r},\omega)$ is the complex induced electron density by a monochromatic perturbing field $E_0 \exp(-i\omega t) \mathbf{e}_x$. The absorption cross section is given by $\sigma_\mathrm{abs} (\omega) = (\omega / c\epsilon_0) \Im \alpha(\omega)$, where
\begin{equation}
\alpha (\omega) = - \frac{e^2}{\hbar \kappa_0} \int x \delta n(\mathbf{r},\omega) d \mathbf{r} \; ,
\label{Eq2}
\end{equation} 
is the dynamical polarizability. Well-converged results are obtained after $2 \times 10^4$ time steps with a total propagation time $T_{\mathrm{max}} \simeq 40~$fs, using a grid spacing of $0.026~$nm.

Optical properties of large isolated sodium clusters whose atoms are arranged in a BCC geometry have been studied extensively in a very recent article by Li \textit{et al.} in the size range $N \le 331$ \cite{Li13}. Our result in \autoref{fig1} for the BCC arrangement is fully consistent with theirs. For an isolated cluster, the atom/TDDFT and jellium/TDDFT optical absorption spectra are qualitatively similar. However, the incorporation of the atomic structure is reflected in an increased linewidth of the main peak for the two atomic structures considered, BCC and ICO (see the upper right panel of \autoref{fig1}). As is well known, this peak corresponds to a dipole LSP or Mie resonance and, under a quantum treatment, its width is due to Landau fragmentation plus non-electronic damping processes \cite{Yannouleas93,Li13}. It is also worth noticing that the resonance frequency for the ICO geometry is blueshifted (around $0.15~$eV) with respect to the BCC. This is mainly due to the overall compression of the ICO arrangement with respect to the sodium bulk ionic density used in the BCC description. Note that the effective diameter of an \textit{unrelaxed} $297$-atom ICO structure with the same mean density than bulk Na would be $2R\simeq2.78$ nm.

%\begin{figure*}[t]
\begin{figure}[htbp]
\begin{center}
\includegraphics[width=\columnwidth]{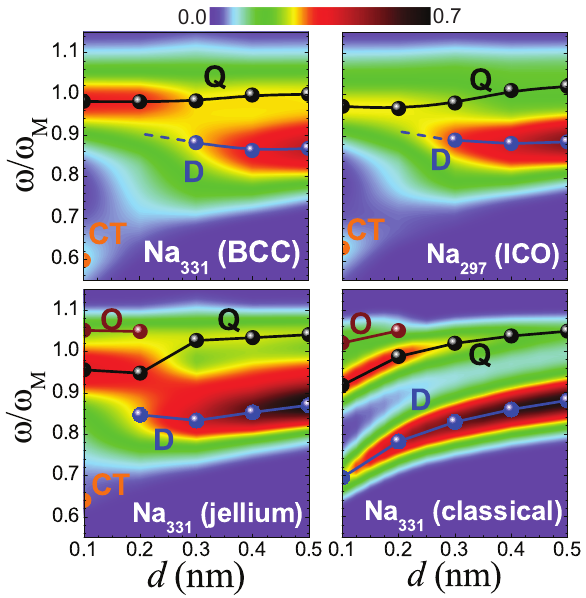}
\caption{(Color online) Contour plots of the normalized absorption cross section [$\sigma_\mathrm{abs}(\omega)/(2\pi R^2)$] of Na cluster dimers vs the distance $d$ as defined in the text. From left to right and top to bottom: BCC, ICO, jellium and classical local optics. The frequencies of the different LSP modes (D: coupled dipole, Q: hybridized quadrupole, O: hybridized octopole, CT: charge transfer) are obtained from the local maxima of the absorption spectrum at each distance.} 
\label{fig2}
\end{center}
%\end{figure*}
\end{figure}

We now analyze the optical absorption of Na cluster dimers in the range of sub-nanometric separations ($0.1 \le d \le 0.5~$nm). The results for the four prescriptions that we are considering (BCC/TDDFT, ICO/TDDFT, jellium/TDDFT and local optics) are presented in \autoref{fig2}. Since the differences between the frequencies of the dipole LSP in the isolated cluster are reflected in the corresponding optical absorption spectra, we normalize the frequency $\omega$ of the incident EM field to the frequency $\omega_\mathrm{M}$ for each prescription. The main trends in the hybridization process can be well understood in terms of classical optics \cite{Nordlander04}. This description breaks down for small separations, where charge transfer between the clusters is possible \cite{Zuloaga09}. As shown in \autoref{fig2}, the jellium/TDDFT and the classical spectra agree well in the range $d \ge 0.4~$nm, where charge transfer is almost negligible. Specifically, under both approximations the dipole mode (D) is red-shifted from the value $\omega_\mathrm{M}$ and a hybridized quadrupole mode (Q) appears at almost identical normalized frequencies. In contrast,  the two atomistic descriptions show a very different behavior. While the normalized frequency of the dipole mode is the same for $d = 0.5~$nm, discrepancies already show up at $d=0.4~$nm. Moreover, the normalized frequency of the mode Q is slightly smaller and the corresponding peak is less resolved in the optical spectra. This can be attributed to the atomic-scale corrugation on the surface of the cluster, affecting the hybridization process which is mediated by the EM near-field.

\begin{figure*}[t]
\begin{center}
\includegraphics[width=\linewidth]{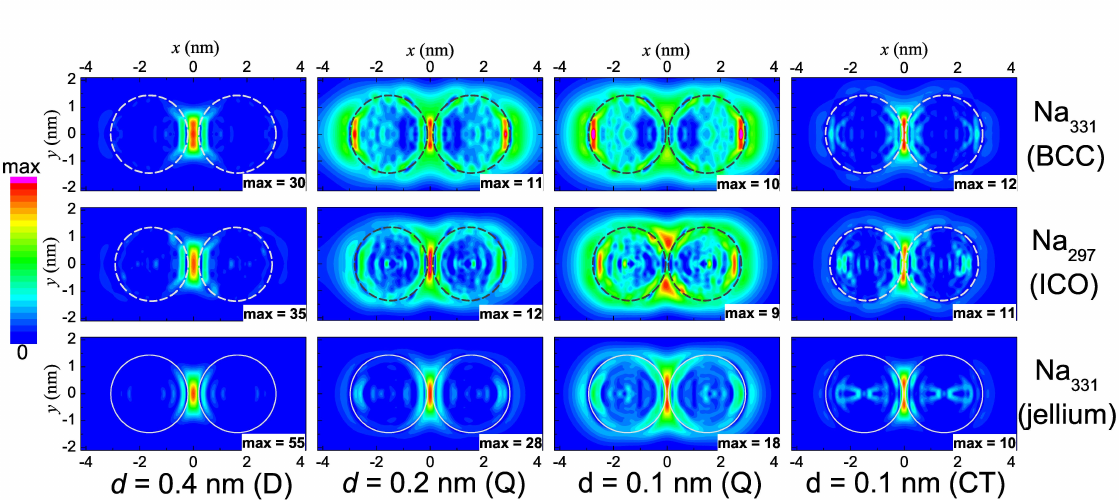}
\caption{(Color online) Contour plots of the electric field amplitude evaluated on the $XY$ plane of the dimer at four selected resonances. Upper row: BCC/TDDFT; middle row: ICO/TDDFT and lower row: jellium/TDDFT calculations. From left to right: dipolar mode D ($d = 0.4~$nm); quadrupole mode Q ($d = 0.2~$nm), Q mode ($d = 0.1~$nm) and charge transfer CT mode ($d = 0.1~$nm). In each panel the E-field amplitude is normalized to its maximum value, which is also shown as a legend.
%Only the positions of the atoms lying in the $XY$ plane are depicted.
}
\label{fig3}
\end{center}
\end{figure*}

As is well-known, the classical description is not valid in the charge-transfer regime. \autoref{fig2} demonstrates that this regime is reached for $d \simeq 0.3~$nm for the jellium model. In this case, the frequency of the D mode stabilizes and the spectral weight of the mode decreases as the nanoparticles get closer. Consequently the hybridized Q mode becomes dominant for $d = 0.2~$nm. Furthermore, a weak hybridized octopole mode O appears. At a distance of $d = 0.1~$nm, the mode D is completely quenched, and a very weak signature of a charge transfer mode (CT) appears instead \cite{Esteban12}. However, the details of the atomic structure in the gap region between the nanoparticles are crucial when determining the intensity of the charge transfer current. In fact, charge transfer effects start to be significant for distances $d = 0.3-0.4~$nm in both BCC and ICO arrangements. In turn, the CT mode in the almost-touching limit ($d = 0.1~$nm) can be clearly discriminated in the absorption spectra when the ionic structure is taken into account. 

\begin{figure}[htbp]
\begin{center}
\includegraphics[width=\columnwidth]{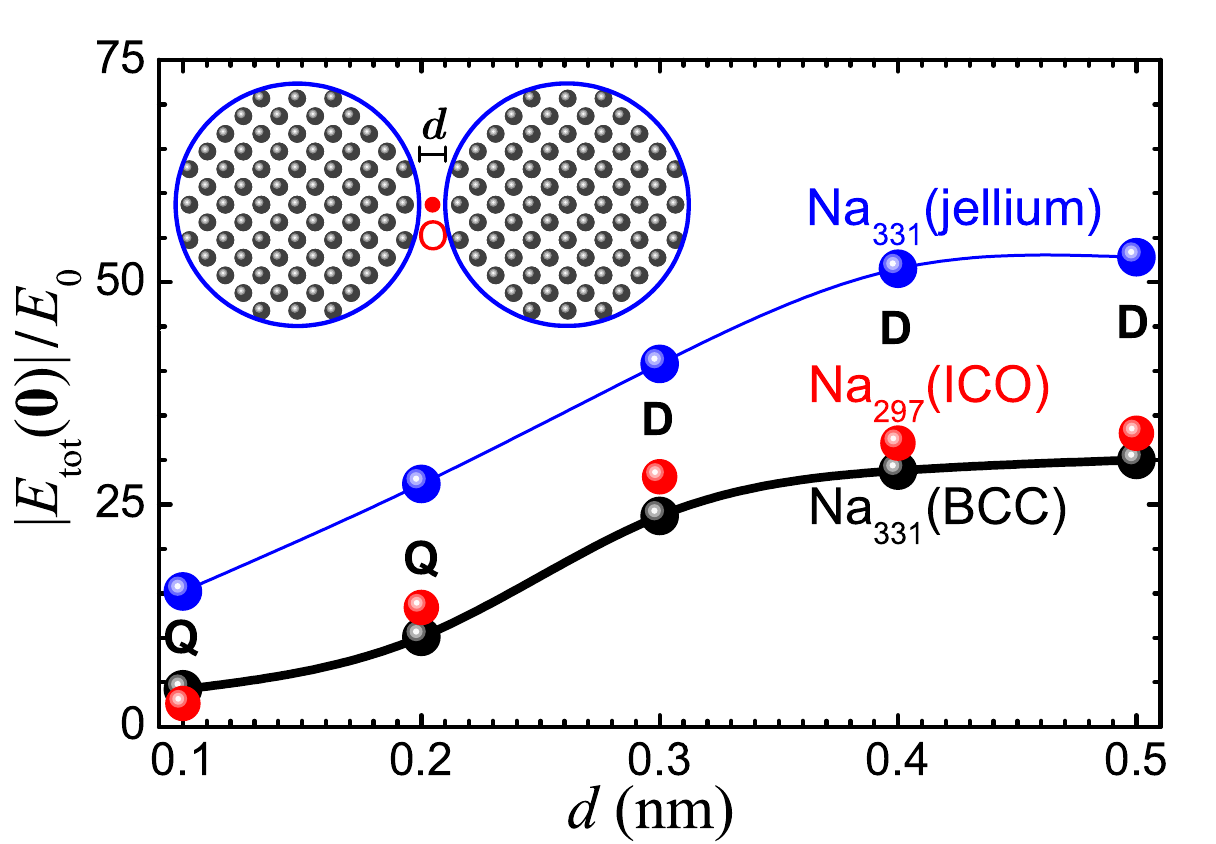}
\caption{(Color online) Electric field enhancement at the center \textbf{O} of the dimer corresponding to the dominant resonance of the absorption spectrum, as indicated in the figure itself, for the three TDDFT calculations: jellium (blue dots), BCC arrangement (black dots) and icosahedral structure (red dots). The connecting lines are just guides to the eye.}
\label{fig4}
\end{center}
\end{figure}

As commented above, a nanoparticle dimer has been considered as a prototypical system to test light harvesting properties of nanoplasmonic devices. Therefore, it is worth analyzing how the atomistic description affects both the modal shape and electric field enhancement (EFE) associated with the different resonances of the cluster dimer. As we have shown in \autoref{fig2}, for $d>0.3~$nm the optical response of the dimer is dominated by the dipolar mode. In the left panels of \autoref{fig3} we render the electric field amplitudes for this D mode calculated at $d=0.4~$nm for BCC/TDDFT (upper panel), ICO/TDDFT (middle panel) and jellium/TDDFT (lower panel) approaches. Notice that the scale is normalized to the maximum value of the E-field amplitude in each case, which is also rendered as a legend. As expected, the modal shape is very similar in the three cases although the D mode for the atom/TDDFT approaches (BCC and ICO) is a bit more delocalized than the jellium/TDDFT counterpart. This results in a smaller EFE at the center of the Na cluster dimer for the two atomistic descriptions: the EFE is reduced by a factor of around $1.5$ when the atomic structure is included in the calculation (see \autoref{fig4}). The influence of the atomic structure in both the modal shape and EFE is even more critical when analyzing the quadrupole mode Q at shorter distances (see central panels of \autoref{fig3}). For the two distances ($d=0.2~$and $d=0.1~$nm), the Q mode for the atom/TDDFT descriptions is much more delocalized than that emerging from the jellium/TDDFT numerical calculation. Whereas in this last case the E-field is still strongly localized in the region between the two clusters, the E-field amplitude for the Q mode in the atomistic calculations is more uniformly distributed along the surfaces of the two clusters. Remarkably, for $d=0.1~$nm, the maximum EFE does not appear at the region between the nanoparticles as in the jellium/TDDFT description, but at the opposite ends of the two clusters.  The right panels of \autoref{fig3} show the modal shape of the CT mode for the three TDDFT calculations. Surprisingly, the CT mode is much less sensitive to the atomic structure than the Q mode and the E-field profiles for the three TDDFT calculations are very similar. 

By using {\it ab-initio} TDDFT calculations for analyzing the optical response of metal nanoparticle dimers, we have been able to demonstrate that the atomic structure of the metal clusters plays a key role for determining accurately both the absorption cross section and electric field enhancement associated with these nanoplasmonic devices. This effect is more critical when the distance between the nanoparticles is smaller than around $0.3~$nm. From a quantitative perspective, the inclusion of the ionic structure into the TDDFT calculations has a similar influence as the incorporation of both the electron density spill-out and non-locality in jellium-based TDDFT calculations when comparing them to the classical EM approaches.  In conclusion, atoms matter in nanoplasmonics and  it is mandatory to go beyond the usual jellium approach in TDDFT calculations if a quantitative description of quantum effects in nanoplasmonic structures is required.

This work has been funded by the European Research Council (ERC-2011-AdG Proposal No. 290981).  We also acknowledge financial support from the European Research Council (ERC-2010-AdG Proposal No. 267374), Spanish Grants (MAT2011-28581-C02-01 and FIS2010-21282-C02), Grupos Consolidados UPV/EHU  (IT-578-13), COST actions: CM1204, MP1306, and EC project CRONOS (Grant number 280879-2). Useful discussions with Prof. J.C. Cuevas and A. Varas are also acknowledged.

\end{document}